# Synthesis of *c*-axis textured CaKFe$_4$As$_4$ superconducting bulk via spark plasma texturing technique

Shigeyuki ISHIDA[a,*], Yoshihisa KAMIYA[b], Yoshinori TSUCHIYA[a], Pavan Kumar Naik SUGALI[a], Yasunori MAWATARI[a], Akira IYO[a], Yoshiyuki YOSHIDA[a], Hiroshi EISAKI[a], Kenji KAWASHIMA[b], Hiraku OGINO[a]

[a]*Research Institute for Advanced Electronics and Photonics, National Institute of Advanced Industrial Science and Technology (AIST), Tsukuba, Ibaraki 305-8568, Japan*

[b]*IMRA JAPAN CO., LTD., Kariya, Aichi 448-8650, Japan*

* Corresponding author.

E-mail: s.ishida@aist.go.jp

**Abstract:** Grain alignment is a key factor that determines the performance of a superconducting bulk. In this study, the spark plasma texturing (SPT) technique was used to fabricate a CaKFe$_4$As$_4$ superconducting bulk. X-ray diffraction and electron backscatter diffraction revealed that the *c*-axes of the CaKFe$_4$As$_4$ grains in the SPT bulk are aligned, demonstrating that the SPT technique is effective in achieving *c*-axis texture. In addition, chemical composition analysis showed that oxide impurities, which affect the grain boundary characteristics that determine the inter-grain critical current density ($J_c$), are randomly distributed in the SPT bulk. Magnetization measurements showed high $J_c$ values of the SPT bulk, reaching 127 kA cm$^{-2}$ and 26 kA cm$^{-2}$ at 4.2 K under a self-field and magnetic field of 5 T, respectively. These results suggest that the SPT technique is a promising approach for obtaining a high-performance superconducting bulk material for high-field applications.

1. **Introduction**

Superconducting materials are utilized in various applications such as high-field magnets, power cables, and electrical devices. In high-field magnet applications, superconductors are typically used in the form of wires, tapes, and bulks. The improved performance through new technologies and materials will lead to an expansion of the operating temperature and magnetic field range, which is crucial for the widespread use of superconductivity applications.

Iron-based superconductors (IBSs) are considered promising candidates for high magnetic field applications because of their large and nearly isotropic upper critical fields [1,2,3]. To date, much effort has been devoted to improving the performance of IBS wires, tapes, and bulks, yielding encouraging results, such as a high critical current density ($J_c$) of 150 kA cm$^{-2}$ at 4.2 K and 10 T [4]. In general, the $J_c$ of the samples in polycrystalline form is limited by the grain boundary (GB) characteristics that determine the inter-grain $J_c$. Therefore, (i) phase purity, (ii) density, and (iii) grain orientation have been recognized as the key factors for achieving a high $J_c$. For IBS wire/tape fabrication, various techniques, such as hot isostatic pressing (HIP), hot/cold pressing, flat/groove rolling, and mechanical alloying, have been used to improve these three factors [5].

Regarding IBS bulk fabrication, previous studies have focused mainly on factors (i) and (ii). For example, high-density $Ba_{1-x}K_xFe_2As_2$ bulks (up to 98% of the theoretical density) were obtained using the HIP technique, and high $J_c$ values of 110 kA cm$^{-2}$ and 12 kA cm$^{-2}$ at 4.2 K under self-field (s.f.) and 5 T, respectively, were achieved [6,7]. Moreover, $J_c$ was improved to 233 kA cm$^{-2}$ (32 kA cm$^{-2}$) at 4.2 K and s.f. (5 T) by eliminating oxide impurities at GBs using a high-performance glovebox and high purity starting materials [8]. We applied another method, namely spark plasma sintering (SPS) (Fig. 1(a)), to obtain a high-density IBS bulk [9]. We succeeded in the fabrication of a high-density (~96%) $CaKFe_4As_4$ bulk using the SPS technique and achieved high $J_c$ values of 81 kA cm$^{-2}$ (18 kA cm$^{-2}$) at 4.2 K and s.f. (5 T). However, the IBS bulks fabricated in these studies exhibited no trace of grain alignment. To further improve the $J_c$ of IBS bulks, it is desirable to investigate the effect of the grain alignment (factor (iii)) in addition to improving the phase purity and density.

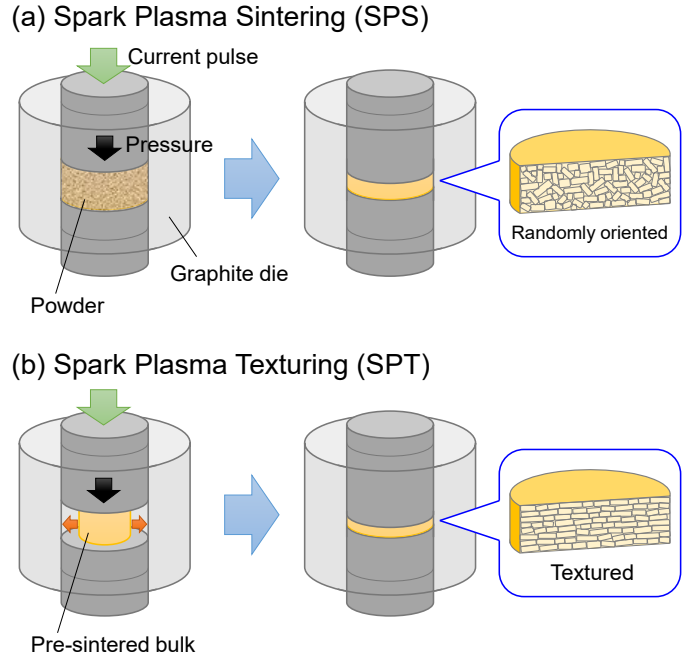

**Fig. 1** Schematic of the process of (a) SPS and (b) SPT, and expected grain orientation in bulk for each process.

Recently, a Ba(Fe$_{1-x}$Co$_x$)$_2$As$_2$ bulk with $c$-axis texture (the $c$-axes of grains are aligned along the pressure applied to pelletize powder) was synthesized using a powder prepared by grinding single crystals [10]. The $J_c$ values of the $c$-axis textured bulk reached 87 kA cm$^{-2}$ (26 kA cm$^{-2}$) at 4.2 K and s.f. (5 T), demonstrating the effectiveness of grain alignment. However, this method requires the growth of large single crystals prior to bulk fabrication, which may become a bottleneck for large bulk fabrication (e.g., 50 mm in diameter) in the future. Here, we focus on a technique called spark plasma texturing (SPT) [11], which is relatively simple and can be applied to large bulk fabrication. SPT is defined as an edge-free SPS technique (Fig. 1(b)), where a pre-sintered sample with sufficient mechanical resistance is placed in a large die. This setup allows for the deformation of the sample perpendicular to the pressure direction, leading to grain alignment. It has been demonstrated that highly textured bulks can be obtained using SPT techniques, such as thermoelectric Ca$_3$Co$_4$O$_9$ [11] and superconducting Bi$_2$Sr$_2$Ca$_2$Cu$_3$O$_{10}$ ceramics [12].

In this study, the SPT technique was used to fabricate a CaKFe$_4$As$_4$ superconducting bulk material. The results revealed the *c*-axis alignment of grains as well as the random distribution of oxide impurities, indicating that deformation during the SPT process notably modifies the microstructure of the bulk. Moreover, the SPT bulk exhibited enhanced $J_c$ values, demonstrating the effectiveness of the SPT technique in improving the $J_c$ performance of IBS bulks.

2. **Experimental procedure**

Polycrystalline CaKFe$_4$As$_4$ powder was synthesized via a solid-state reaction. Details of the powder synthesis procedure are described in a previous work [9]. CaKFe$_4$As$_4$ powder (~2 g) was inserted into a graphite die with an inner diameter of 10 mm and placed in an SPS apparatus (SS Alloy Co., Ltd.). A uniaxial load of 4 kN (corresponding to a pressure of ~50 MPa for a 10 mm diameter die) was applied to the graphite die. The sample chamber was evacuated and filled with argon gas to the pressure of 0.5 atm. The graphite die was heated to 750 °C for sintering at a rate of 50 °C min$^{-1}$ by applying current pulses, held for ~10 min until densification settled, and then furnace-cooled to room temperature. The sintered bulk was named "SPS bulk." The SPS bulk had a diameter of ~10 mm, thickness of ~5.2 mm, and density of 5.11(1) g cm$^{-3}$ (corresponding to 97.9(2)% of the theoretical density of CaKFe$_4$As$_4$ (5.22 g cm$^{-3}$ [13])). The SPS bulk was thereafter inserted into a larger graphite die with an inner diameter of 20 mm and placed in the SPS apparatus. The sample chamber was purged and filled with argon gas at 0.5 atm. A relatively small uniaxial pressure of 1 kN was applied to the graphite die to avoid cracking the SPS bulk. The graphite die was heated to 750 °C at a rate of 50 °C min$^{-1}$, where the bulk deformation occurred. Subsequently, the uniaxial load was gradually increased to 16 kN. This load corresponds to a pressure of ~50 MPa for a 20 mm diameter die, which is the same pressure applied to the SPS bulk. Thus, we obtained a bulk with diameter of ~20 mm, thickness of ~1.2 mm, and density of 5.01(1) g cm$^{-3}$ (95.9(2)%), which was named "SPT bulk".

Room temperature X-ray diffraction measurements were performed using Cu Kα radiation (Rigaku, Ultima IV) with a high-speed detection system (Rigaku, D/teX Ultra). The microstructure and chemical composition were examined using a field emission scanning electron microscope (Carl Zeiss, ULTRA55) equipped with an energy dispersive X-ray spectrometer (Thermo Fisher Scientific, NSS312E). Prior to the microstructure observations, the bulk samples were cut in half near the center, and the cross sections were polished by Ar ion milling using a cross-sectional polisher (Gatan, PECS II Model 685). The grain orientation was analyzed using electron backscatter diffraction (EBSD) detector (Oxford Instruments, Symmetry) on the cross section with a step size of 0.2 mm. The clean-up procedure of the EBSD maps was performed by the standard wild-spike correction method of the Aztec software (Oxford Instruments). The minimum grain size and minimum misorientation in the present EBSD analysis were defined to be 0.5 μm in diameter and 5 degree, respectively. Magnetization measurements were performed on small rectangular pieces (~2 × 1.5 × 1 mm$^3$) using a SQUID magnetometer (Quantum Design, MPMS). The rectangular pieces were cut from near the center of the remaining half of the bulk samples so that the results of magnetization measurements and microstructure observations could be compared. Magnetization hysteresis loops were measured up to a magnetic field of 7 T at temperatures of 4.2 K, 10 K, 15 K, 20 K, 25 K, and 30 K. The critical current density ($J_c$) was calculated based on Bean's critical state model [14]: $J_c = 20\Delta M/w(1 - w/3l)$, where the unit of $J_c$ is A cm$^{-2}$, $\Delta M$ (in emu cm$^{-3}$) is the width of the magnetization hysteresis loops in emu, and $l$ and $w$ ($l > w$) are the dimensions of the samples in cm.

3. Results and discussion
3. 1 X-ray diffraction (XRD) patterns
Figure 2 shows the XRD patterns of the CaKFe$_4$As$_4$ powder (black), SPS bulk surface (blue), and SPT bulk surface (red). In all the samples, CaKFe$_4$As$_4$ (*P*4/*mmm*, SG123) was the main phase (several peaks were indexed), with small amounts of CaFe$_2$As$_2$ and FeAs as the secondary phases. According to the XRD patterns, no significant decomposition of the CaKFe$_4$As$_4$ phase occurred during the SPS and SPT processes.

Next, we focused on the peak intensities to examine the degree of $c$-axis alignment in bulk samples. The intensity of the (103) peak was the strongest for the powder and SPS bulk. The intensity of the (002) peak ((200) peak) was slightly enhanced (suppressed) in the SPS bulk, suggesting that the $c$-axes of the grains were only weakly aligned along the pressure direction. However, for the SPT bulk, the (002) peak intensity becomes stronger than that of the (103) peak. Moreover, the (200) and (110) peaks almost disappear. These results indicate that the $c$-axes of the grains are well aligned in the SPT bulk.

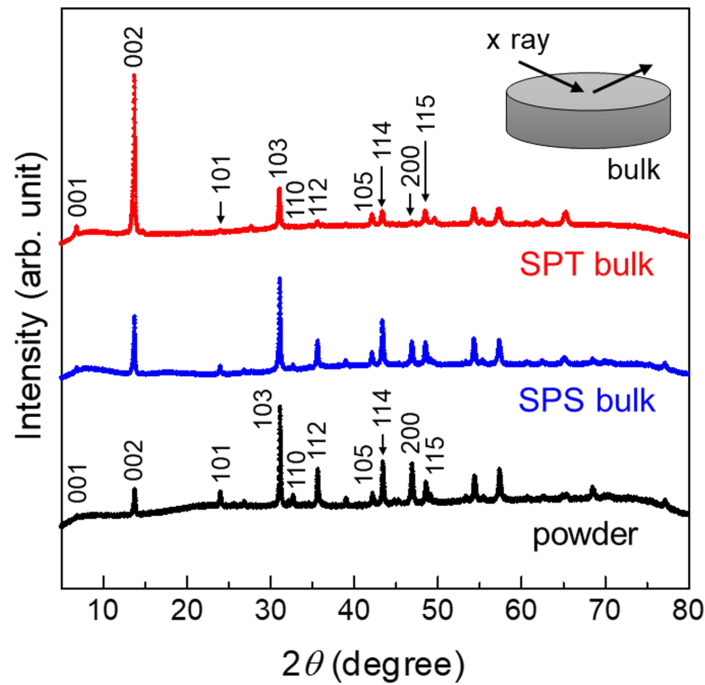

**Fig. 2** X-ray diffraction patterns of CaKFe$_4$As$_4$ powder (black), SPS bulk (blue), and SPT bulk (red). Measurements for bulks were performed on a surface perpendicular to the pressure direction, as shown in inset.

To quantify the degree of $c$-axis alignment in these bulk samples, the orientation factor ($F$) was evaluated based on the XRD patterns using the Lotgering method [15]: $F = (\rho - \rho_0)/(1 - \rho_0)$, where $\rho = \Sigma I(00l)/\Sigma I(hkl)$ and $\rho_0 = \Sigma I_0(00l)/\Sigma I_0(hkl)$, where $I(hkl)$ and $I_0(hkl)$ are the intensities of the $hkl$ peaks for bulk and powder (randomly oriented) samples, respectively. The estimated $F$ values are 0.10(1) and 0.54(1) for the SPS bulk and SPT bulk, respectively. The $F$ value of SPT bulk was comparable with those reported for IBS tapes [16] and *c*-axis textured Ba(Fe$_{1-x}$Co$_x$)$_2$As$_2$ bulk [10]. Thus, the XRD pattern of the SPT bulk surface indicates that the *c*-axis texture was successfully obtained in the SPT bulk.

## 3. 2  Grain orientation mapping

Since the degree of *c*-axis alignment could be different between the bulk surface and the interior, the microstructures inside the SPS and SPT bulks were investigated using scanning electron microscopy (SEM) of the cross section, as schematically shown in Fig. 3(a).

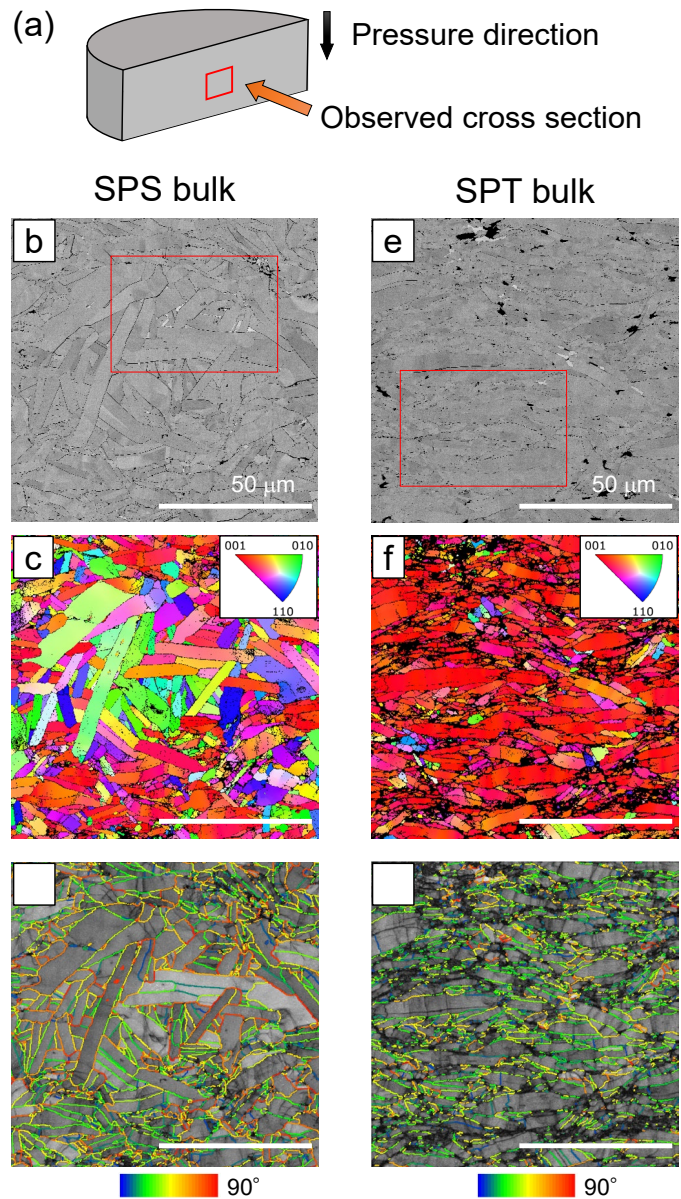

**Fig. 3** Grain orientation analysis performed on cross section schematically shown in (a). (b) SEM image and (c) EBSD orientation map colored with inverse pole figure scheme shown in inset for SPS bulk. (d) Misorientation angle across the grain boundaries for SPS bulk. Hot (cold) colors indicate large (small) angle variations. (e)-(g) Same data set for SPT bulk. The white bars in (b)-(g) indicate 50 μm.

Figure 3(b) shows the SEM image of the SPS bulk cross section. Rectangular grains with long sides of several tens of micrometers and short sides of several micrometers were oriented in random directions. Considering the layered crystal structure of $CaKFe_4As_4$, the long side is expected to correspond to the *ab*-plane direction, and the short side to the *c*-axis direction. The random orientation of the grains is consistent with the XRD pattern of the SPS bulk surface, showing no trace of *c*-axis texture. Figure 3(c) shows the electron backscatter diffraction (EBSD) orientation map along the vertical direction (pressure direction), colored with the inverse pole figure scheme shown in the inset. The grains exhibited various colors, indicating that the crystal axes were randomly oriented. The results confirm the absence of *c*-axis texture in bulk SPS. In addition, the misorientation angle across the grain boundaries is plotted in Fig. 3(d). Large misorientation angles (hot colored GBs) can be found reflecting the random orientation of the grains.

Figure 3(e) shows the SEM image of the SPT bulk cross section. In contrast with those of the SPS bulk (Fig. 3(b)), the long sides of the grains are predominantly oriented in the horizontal direction. The EBSD orientation map (Fig. 3(f)) indicates that most of the grains are colored red, that is, the dominant orientation of the grains is (001) along the pressure direction, confirming the *c*-axis texture inside the SPT bulk. As a result, the misorientation angles in the SPT bulk become smaller than those in the SPS bulk as shown in Fig. 3(g) where most of the GBs are cold-colored. These results demonstrate that SPT is an effective approach for achieving *c*-axis texture, which is one of the key factors for improving $J_c$ performance in IBS bulks.

### 3. 3  Energy dispersive X-ray spectroscopy (EDS) analysis

A previous study of CaKFe$_4$As$_4$ SPS bulk [9] revealed the existence of oxide impurities in GBs, which have been considered a limiting factor for $J_c$ performance. The oxygen distributions in the SPS and SPT bulks were investigated using EDS.

Figure 4(a) shows an enlarged SEM image of the SPS bulk cross section, indicated by the red square in Fig. 3(b). Based on chemical composition analysis, the bright regions were identified as FeAs impurities. It can be seen that some CaKFe$_4$As$_4$ grains are surrounded by dark lines, making GBs recognizable. Figure 4(b) shows oxygen mapping in the same area. The dark lines in Fig. 4(a) were identified as oxide impurities distributed along the GBs. Also, there were some distributions of oxygen within the grains. These results indicate that the as-synthesized CaKFe$_4$As$_4$ grains were covered with oxide impurities and also contained oxide inclusions, possibly arising from the oxidation of the starting reagents and residual oxygen in the glove box.

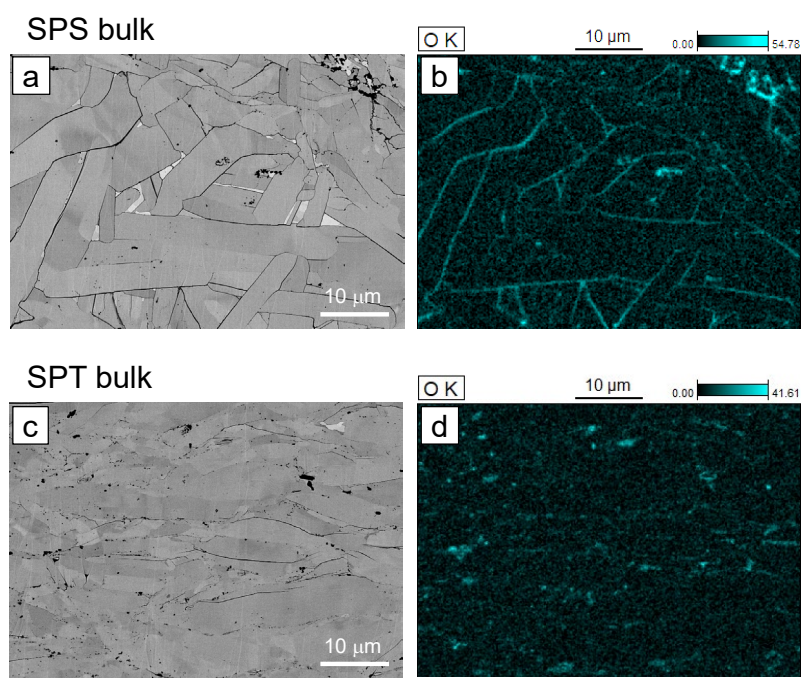

**Fig. 4** (a) SEM image and (b) oxygen mapping for SPS bulk in area indicated by red square in Figure 3(b). (c), (d) Same data set for SPT bulk of red square area in Figure 3(e).

Figures 4(c) and (d) show the SEM image and the oxygen mapping of the SPT bulk, respectively. In contrast with the case of SPS bulk, the GBs are less visible (Fig. 4(c)), and the oxide impurities show a random distribution (Fig. 4(d)) in the SPT bulk. As shown in Figs. 3(d) and (g), the SPT bulk shows relatively smaller misorientation angles compared with the SPS bulk. In general, the degree of grain misorientation affects the visibility of the grains in the SEM images. The small misorientation angles may be a reason for the less-visible GBs in the SPT bulk. Also, in the case of SPS bulk, the oxide impurities surrounding the grains made GBs clear. On the other hand, in the SPT bulk, the distribution of oxide impurities became random possibly due to the deformation and reorientation of grains during the SPT process, which may also result in the less-visible grain boundaries. Thus, the SPT process aligns the grain orientation as well as changes the distribution of oxide impurities in the bulk.

### 3. 4  Critical current properties

The $J_c$ properties of the SPT bulk were evaluated based on the magnetization measurements. Figure 5(a) shows the magnetic field ($H$) dependence of $J_c$ ($J_c - H$) at several temperatures, where $H$ is applied along the pressure direction. Here, the supercurrent flows perpendicular to the pressure direction (along the horizontal direction in Fig. 3(d)), that is, mainly along the $ab$-plane direction. The $J_c$ values of the SPT bulk are 127 kA cm$^{-2}$ and 26 kA cm$^{-2}$ at 4.2 K under s.f. and 5 T, respectively, which are higher than those of our SPS bulk [9], and comparable with Ba$_{1-x}$K$_x$Fe$_2$As$_2$ bulk fabricated by HIP [8] and $c$-axis textured Ba(Fe$_{1-x}$Co$_x$)$_2$As$_2$ bulk [10]. Regarding $H$ dependence, the second magnetization peak (SMP), which is often observed in various IBS single crystals including CaKFe$_4$As$_4$ [17,18,19,20], is absent in the SPT bulk. Note that SMP has been observed for CaKFe$_4$As$_4$ bulks synthesized without SPS [21,22], which could be associated with the non-negligible contribution from intra-grain $J_c$. Then, $J_c - H$ of the SPT bulk is considered to reflect the inter-grain $J_c$ contribution.

Next, $J_c$ under $H$ perpendicular to the pressure direction was evaluated (the measurement configurations are shown in the inset of Fig. 5(b)). In this case, the supercurrent flows perpendicular to the pressure direction and along the pressure direction (i.e., along the $c$-axis direction). Although there are two $J_c$ contributions, for simplicity, the average $J_c$ was calculated assuming a uniform current

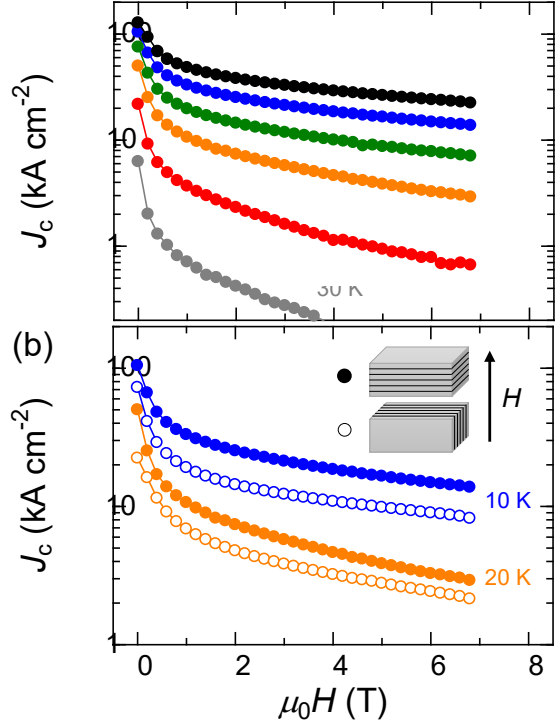

**Fig. 5** (a) Magnetic field dependence of $J_c$ ($J_c$ – $H$) for SPT bulk under $H$ parallel to pressure direction. (b) Comparison of $J_c$ – $H$ at 10 K (blue) and 20 K (orange) under $H$ applied parallel (solid circles) and perpendicular (open circles) to the pressure direction, respectively. Inset shows two measurement configurations.

flow. As shown in Fig. 5(b), the average $J_c$ in this configuration (open circles) was smaller than that under $H$ along the pressure direction (solid circles). This indicates that $J_c$ along the pressure direction (along the $c$-axis) was smaller than that perpendicular to the pressure direction (along the $ab$-plane). Considering that the inter-grain $J_c$ is generally dominated by GB characteristics, the results indicate that in the $c$-axis textured IBS bulk, the GB characteristics between horizontally aligned grains and those between stacked grains are significantly different. In the actual use of superconducting bulks, a disk-shaped bulk material is expected to trap magnetic fields along the axial direction. The supercurrent flowing perpendicular to the axial direction, that is, the higher $J_c$ in the present SPT bulk, is relevant in determining the performance of the bulk. Thus, the present $c$-axis texture is effective for improving the performance of IBS bulks.

## Conclusions

In summary, a $c$-axis textured CaKFe$_4$As$_4$ bulk was successfully obtained using SPT. It was also found that deformation during the SPT process likely affected the distribution of the oxide impurities. The grain alignment and modified oxide impurity distribution are expected to be advantageous for obtaining better GB properties. Indeed, the SPT bulk showed high $J_c$ values reaching 127 kA cm$^{-2}$ (26 kA cm$^{-2}$) at 4.2 K and s.f. (5 T), which are comparable with the highest values reported for IBS bulks. Thus, the SPT technique is a promising approach for realizing high-performance IBS bulk materials. Considering the degree of $c$-axis alignment ($F$ = 0.54), as well as the non-negligible amount of oxide impurities in the present SPT bulk, there is much room for further improvement of $J_c$ by optimizing the SPT process and reducing oxygen contamination.

## Acknowledgements

This work was supported by the Japan Society for the Promotion of Science (JSPS) Grant-in-Aid for Scientific Research on Innovative Areas and KAKENHI (JSPS Grant Numbers 16H06439 and 19H02179). SEM/EDS measurements and EBSD analysis were performed at JFE Techno-Research Corporation with technical support from Dr. S. Tsukimoto.

## Competing interest

The authors have no competing interests to declare that are relevant to the content of this article.